\documentclass[12pt,cite,epsf,epsfig]{article}
\usepackage{epsfig}

\begin{document}

\author{S. DEV\thanks{dev5703@yahoo.com} and
SANJEEV KUMAR\thanks{sanjeev3kumar@yahoo.co.in}}
\title{Neutrino Parameter Space for a Vanishing $ee$ Element in the Neutrino
Mass Matrix}
\date{Department of Physics, Himachal Pradesh University, Shimla 171005, INDIA}
\maketitle

\begin{abstract}
The consequences of a texture zero at the $ee$ entry of neutrino
mass matrix in the flavor basis, which also implies a vanishing
effective Majorana mass for neutrinoless double beta decay, have
been studied for Majorana neutrinos. The neutrino parameter space
under this condition has been constrained in the light of all
available neutrino data including the CHOOZ bound on $s_{13}^{2}$.
\end{abstract}

One of the fundamental goals of neutrino physics is to determine
the neutrino mass matrix given by
\begin{equation}
(m_{\nu})_{\alpha
\beta}=(Um_{\nu}^{diag}U^{T})_{\alpha\beta};\hspace{10pt}
\alpha,\beta=e,\mu,\tau
\end{equation}
for Majorana neutrinos where $m_{\nu}^{diag}=\{m_1,m_2,m_3\}$.
Since $m_{\nu}$ is a symmetric matrix, it is specified by nine
physical parameters viz. the three neutrino masses ($m_1$, $m_2$,
$ m_3$), three mixing angles ($\theta _{12}$, $\theta _{23}$,
$\theta_{13}$), the Dirac-type CP-violating phase ($\delta $), and
two Majorana-type CP-violating phases ($\alpha $, $\beta $). Out
of these nine parameters, only the first six have been constrained
to a reasonable degree of precision but the Dirac-type
CP-violating phase and the two Majorana-type CP-violating phases
remain unconstrained at present. While the Dirac-type CP-violating
phase is expected to be constrained from the study of CP-violation
in long baseline neutrino oscillations, the two Majorana-type
CP-violating phases can, hopefully, be constrained from
neutrinoless double $\beta$ decay ($0\nu\beta\beta$).

The rate for $0\nu\beta\beta$ decay is proportional to the
effective mass defined as
\begin{equation}
\left|m_{ee}\right|=\left|\sum_{i} m_i U_{ei} \right|^2
\end{equation}
which, in fact, is the magnitude of the first element of the
neutrino mass matrix in the charged lepton flavor basis. Thus, the
$0\nu\beta\beta$ \cite{1} decay provides us an unique opportunity
to probe directly one of the elements of the neutrino mass matrix.
The non-observation of $0\nu\beta\beta$ decay constrains the
effective mass $|m_{ee}|$ to be close to zero. The possibility
$m_{ee}=0$ has been examined in the literature \cite{2,3,4,5,6,7}
earlier under certain special conditions. However, a general study
of the neutrino parameter space for $m_{ee}=0$ with the full
interplay of the Majorana phases is still lacking.

In the present work, we examine the consequences of the vanishing
effective Majorana mass for the neutrino parameter space. A study
of this condition has important implications for the texture
specific phenomenological neutrino mass matrices. In particular,
the phenomenological conclusions obtained from the condition of
vanishing effective Majorana mass will be valid for all texture
zero schemes in the charged lepton basis in which the first
element is zero.

The neutrino mixing matrix i.e. the PMNS matrix $U$ can be
parameterized in terms of three mixing angles ($\theta _{12}$,
$\theta _{23}$, $\theta _{13}$), the Dirac CP-violating phase
($\delta $), and the two Majorana CP-violating phases ($\alpha $,
$\beta $) in the following manner \cite{8}:
\begin{equation}
U=\left(
\begin{array}{ccc}
c_{12}c_{13} & s_{12}c_{13} & s_{13}e^{-i\delta } \\
-s_{12}c_{23}-c_{12}s_{23}s_{13}e^{i\delta } &
c_{12}c_{23}-s_{12}s_{23}s_{13}e^{i\delta } & s_{23}c_{13} \\
s_{12}s_{23}-c_{12}c_{23}s_{13}e^{i\delta } &
-c_{12}s_{23}-s_{12}c_{23}s_{13}e^{i\delta } & c_{23}c_{13}
\end{array}
\right) \left(
\begin{array}{ccc}
1 & 0 & 0 \\
0 & e^{i\alpha } & 0 \\
0 & 0 & e^{i\left( \beta +\delta \right) }
\end{array}
\right)
\end{equation}
where $c_{ij}=\cos \theta _{ij}$ and $s_{ij}=\sin \theta _{ij}$.
This parameterization has the advantage that  the Dirac
CP-violating phase $\delta$ is formally absent in the expression
of the effective Majorana mass $m_{ee}$ probed in $0\nu \beta
\beta $ decays. The neutrino mass matrix is a complex symmetric
matrix described by nine parameters as discussed above. For the
above parameterisation, the effective Majorana mass is given by
\begin{equation}
m_{ee}=c_{13}^{2}c_{12}^{2}m_{1}+c_{13}^{2}s_{12}^{2}m_{2}e^{2i\alpha
}+s_{13}^{2}m_{3}e^{2i\beta }.
\end{equation}
The current experimental bound on the effective Majorana mass is
\cite{9,10}
\begin{equation}
\left| m_{ee}\right| \leq 0.35\zeta eV
\end{equation}
where the uncertainty in the calculation of the nuclear matrix
element of $0\nu \beta \beta $ is contained in the factor $\zeta
=O(1)$. The element $m_{ee}$ of neutrino mass matrix depends on
seven (out of a total of nine) parameters viz. $m_{1}$, $m_{2}$,
$m_{3}$, $\theta _{12}$, $\theta _{13}$, $\alpha $ and $\beta $.
The mixing angle $\theta _{12}$ is known from the solar and
KamLAND neutrino data. The masses $m_{2}$ and $m_{3}$ can be
calculated from the two mass-squared differences $\Delta
m_{12}^{2}$ and $\Delta m_{23}^{2}$ which has been measured
experimentally in the solar and atmospheric neutrino experiments.
For normal hierarchy (NH), $ m_{1}$ is the lightest neutrino mass
and the masses $m_2$ and $m_3$ can be expressed in terms of $m_1$
in the following way
\begin{equation}
m_{2}=\sqrt{m_{1}^{2}+\Delta m_{12}^{2}}
\end{equation}
and
\begin{equation}
m_{3}=\sqrt{m_{2}^{2}+\Delta m_{23}^{2}}.
\end{equation}
For inverted hierarchy (IH), $m_{3}$ is the lightest neutrino mass
and the other two masses are given by
\begin{equation}
m_{1}=\sqrt{m_{3}^{2}-\Delta m_{23}^{2}}
\end{equation}
and
\begin{equation}
m_{2}=\sqrt{m_{1}^{2}+\Delta m_{12}^{2}}.
\end{equation}
Thus, two neutrino masses are known from the solar and atmospheric
mass-squared differences viz. $\Delta m^2_{12}$ and $\Delta
m^2_{23}$ in terms of the lightest neutrino mass $m_1$ for NH and
$m_3$ for IH.

The best fit, 1 and 3 sigma values of the oscillation parameters
are \cite {11}:
\begin{eqnarray}
\Delta m_{12}^{2} &=&7.9_{-0.3,0.8}^{+0.3,1.0}\times
10^{-5}eV^{2}, \nonumber \\ s_{12}^{2}
&=&0.31_{-0.03,0.07}^{+0.02,0.09},  \nonumber \\ \Delta m_{23}^{2}
&=&\pm 2.2_{-0.27,0.8}^{+0.37,1.1}\times 10^{-3}eV^{2}, \nonumber
\\ s_{23}^{2} &=&0.50_{-0.05,0.16}^{+0.06,0.18},  \nonumber \\
s_{13}^{2} &<&0.012(0.046).
\end{eqnarray}
The best-fit value of $s^2_{13}$ is zero. The positive and
negative signs of $\Delta m^2$ correspond to the normal and
inverted hierarchy, respectively.

Thus, the element $m_{ee}$ is now a function of four unknown
parameters viz. the mixing angle $\theta _{13} $ and the two
Majorana CP-violating phases ($\alpha $, $\beta $) and the
lightest neutrino mass ($m_{1}$ for NH and $m_{3}$ for IH). The
study of the element $m_{ee}$ is, therefore, important in order to
decode information about these unknown parameters.

For $s^2_{13}=0$, the effective Majorana mass $m_{ee}$ is given by
\begin{equation}
m_{ee}=c^2_{12}m_1+s^2_{12}m_2 e^{2i \alpha}
\end{equation}
which vanishes if
\begin{equation}
\alpha=\left(n+\frac{1}{2}\right) \pi
\end{equation}
and
\begin{equation}
\frac{m_1}{m_2}=\frac{s^2_{12}}{c^2_{12}}.
\end{equation}
Substituting the value of $m_2$ from Eq. (6) or Eq. (9) in Eq.
(13), we obtain
\begin{equation}
m_1= s^2_{12}\sqrt{\frac{\Delta m^2_{12}}{\cos 2 \theta_{12}}}
\end{equation}
both for normal and inverted hierarchies. Thus, there exists a
point on the ($\alpha,m_1$) plane at which $m_{ee}$ vanishes for
$s_{13}^2=0$ for both the hierarchies which is a consequence of
the fact that the effective Majorana mass $m_{ee}$ is independent
of $m_3$ and, hence, independent of the hierarchy for
$s_{13}^2=0$. This case has been examined earlier \cite{2}.

Now we examine the consequences of a vanishing effective Majorana
mass for non-zero $s_{13}^2$. The element $m_{ee}$ depends on the
four parameters viz. the lightest neutrino mass ($m_{1}$ for NH
and $m_{3}$ for IH), the mixing angle $\theta _{13}$ and the two
Majorana CP-violating phases $\alpha $ and $\beta $. Now, we
demand that $m_{ee}=0$ which requires that both the real and
imaginary parts of $m_{ee}$ vanish which yields two conditions on
the neutrino parameter space viz.
\begin{equation}
Re(m_{ee})=c_{13}^{2}c_{12}^{2}m_{1}+c_{13}^{2}s_{12}^{2}m_{2}
\cos 2\alpha +s_{13}^{2}m_{3} \cos 2\beta=0
\end{equation}
and
\begin{equation}
Im(m_{ee})=c_{13}^{2}s_{12}^{2}m_{2} \sin2\alpha
+s_{13}^{2}m_{3}\sin2\beta=0
\end{equation}
so that we can express any two variables in terms of the remaining
two variables. In fact, the condition $m_{ee}=0$ is more
restrictive than the actual measurement of this element from
$0\nu\beta\beta$ decay which yields only one constraint on the
above four-dimensional neutrino parameter space. In the following,
we express $\beta$ and $s_{13}$ in terms of $ m_{1}$ and $\alpha$
to examine the neutrino parameter space allowed by the condition
$m_{ee}=0$ on the ($\alpha ,m_{1}$) plane. We have, already,
obtained a point on the ($\alpha ,m_{1}$) plane corresponding to
$s_{13}^2=0$ and arbitrary values of $\beta$ [cf. Eqs. (12,14)].
Next, we examine the whole region for $m_{ee}=0$ about this point
on the ($\alpha,m_{1}$) plane for different values of $\beta$ and
for $s_{13}^2$ within the CHOOZ bound.

From Eqs. (15) and (16), we obtain
\begin{equation}
s_{13}^{2}=\frac{{m_{1}c_{12}^{2}+m_{2}s_{12}^{2}\cos 2\alpha
}}{{m_{1}c_{12}^{2}+m_{2}s_{12}^{2}\cos 2\alpha }-{m_{3}\cos
2\beta}}
\end{equation}
and
\begin{equation}
s_{13}^{2}=\frac{{m_{2}s_{12}^{2}\sin 2\alpha
}}{{m_{2}s_{12}^{2}\sin 2\alpha }-{m_{3}\sin 2\beta }}
\end{equation}
respectively.

From Eqs. (17) and (18), one can obtain the mass ratios
\begin{equation}
\frac{m_{1}}{m_{2}}=\frac{s_{12}^{2}\sin 2\left( \alpha -\beta \right)}{%
c_{12}^{2}\sin 2\beta }
\end{equation}
and
\begin{equation}
\frac{m_{2}}{m_{3}}=-\frac{s_{13}^{2}\sin 2\beta }{c_{13}^{2}s_{12}^{2}\sin
2\alpha }.
\end{equation}
For $m_{ee}=0$ to hold, Eqs. (17) and (18) must hold
simultaneously which implies
\begin{equation}
\sin 2\beta =\pm \frac{s_{12}^{2}m_{2}}{M}\sin 2\alpha
\end{equation}
where
\begin{equation}
M=\sqrt{%
m_{1}^{2}c_{12}^{4}+m_{2}^{2}s_{12}^{4}+2m_{1}m_{2}s_{12}^{2}c_{12}^{2}\cos
2\alpha }.
\end{equation}
and $s_{13}$ is given by
\begin{equation}
s_{13}^{2}=\frac{M}{M\mp m_{3}}.
\end{equation}
where $\mp$ signs in front of $m_{3}$ correspond to the two signs
of $\sin 2\beta $ in Eq. (21), respectively. The physical
requirement $s_{13}^{2}<1$ implies that $m_{ee}=0$ if
\begin{equation}
\sin 2\beta =-\frac{s_{12}^{2}m_{2}}{M}\sin 2\alpha,
\end{equation}
and
\begin{equation}
s_{13}^{2}=\frac{M}{M+m_{3}}.
\end{equation}
It is interesting to note that $\beta$ becomes indeterminate when
$m_1/m_2=s_{12}^2/c_{12}^2$ and $\alpha=(n+\frac{1}{2})\pi$ for
$s_{13}^2=0$ since $M=0$ at this point. This is consistent with
our earlier result that $m_{ee}$ can vanish for arbitrary values
of $\beta$ when $s_{13}^2=0$.

Substituting the value of $\beta $ from Eq. (24) in Eq. (20), we
obtain
\begin{equation}
m_3=\frac{c^2_{13} M}{s_{13}^2}.
\end{equation}
Note that R.H.S. of Eq. (26) becomes indeterminate at the point
$m_1/m_2=s_{12}^2/c_{12}^2$ for $s_{13}^2=0$ since $M=0$ at this
point which is consistent with our earlier remark that $m_{ee}$
can vanish for $s_{13}^2=0$ irrespective of the values of $m_3$.
One can solve Eq. (26) for $\alpha$ to obtain
\begin{equation}
\cos 2\alpha =\frac{
m_{3}^{2}s_{13}^{4}-c_{13}^{4}(m_{1}^{2}c_{12}^{4}+m_{2}^{2}s_{12}^{4})}{
2m_{1}m_{2}s_{12}^{2}c_{12}^{2}c_{13}^{4}}
\end{equation}
which corresponds to
\begin{equation}
\cos 2\beta =-\frac{
m_{3}^{2}s_{13}^{4}+c_{13}^{4}(m_{2}^{2}s_{12}^{4}-m_{1}^{2}c_{12}^{4})}{
2m_{2}m_{3}s_{12}^{2}s_{13}^{2}c_{13}^{2}}.
\end{equation}

\begin{figure}[tb]
\begin{center}
\rotatebox{0}{\epsfig{file=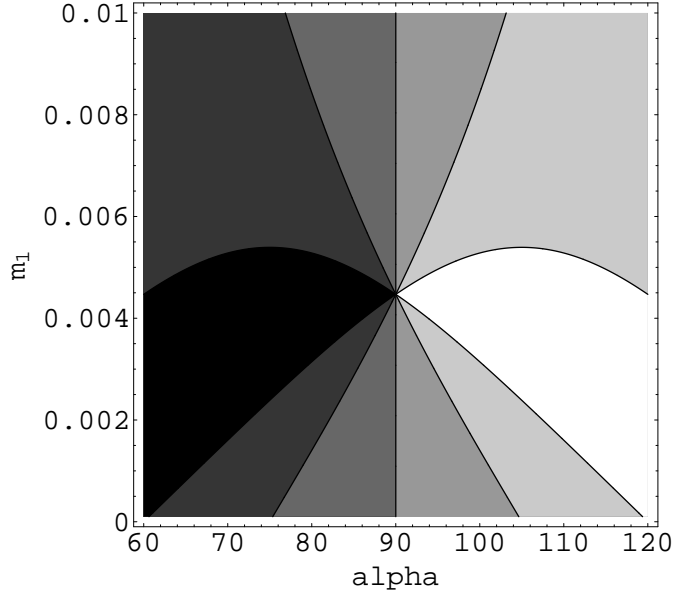, width=9.0cm, height=9.0cm}}
\end{center}
\caption{The contours of constant $\sin 2\beta $ on
($\alpha,m_{1}$) plane as demanded by $m_{ee} =0$. The central
vertical line is for $\sin 2\beta =0$. Immediately right (left) to
it is the line $\sin 2\beta =\frac{1}{2}(-\frac{1}{2})$ followed
by the line for $\sin 2\beta
=\frac{\sqrt{3}}{2}(-\frac{\sqrt{3}}{2})$.}
\end{figure}

Eqs. (24) and (25) can be used to examine the allowed
($\alpha,m_1$) parameter space for $m_{ee}=0$. Note that $\sin 2
\alpha$ and  $\sin 2 \beta$ should have opposite signs. A natural
choice is to take $\alpha$ centered around $90^o$ (corresponding
to $n=0$ in Eq. (12)) and $\beta$ centered around $0^o$. With this
particular choice, $\beta$ should lie in the fourth quadrant (i.e.
$\beta<0^o$), if $\alpha$ lies in the first quadrant (i.e. $
\alpha<90^o$) and $\beta$ should lie in the first quadrant (i.e.
$\beta>0^o$), if $\alpha$ lies in the second quadrant (i.e. $
\alpha>90^o$). In other words, $\alpha$ should be ahead of $\beta$
by one quadrant. This is apparent in Fig. 1 where we have plotted
the contours of constant $\beta$ on the ($\alpha,m_1$) plane. The
central vertical line corresponds to $\alpha=90^o$ and
$\beta=0^o$. The left half of the plot corresponds to $\alpha
<90^{o}$ and $\beta <0^o$ whereas the right half of the plot corresponds to
$\alpha>90^{o}$ and $\beta>0^o$. All the curves corresponding to
different values of $\beta$ intersect at a certain point on the
central vertical line. This point has been obtained earlier while
discussing the $s_{13}^2=0$ case. At this point, $m_{ee}=0$
irrespective of the value of $\beta$. For $\alpha=90^o$, Eq. (25)
gives
\begin{equation}
s_{13}^{2}=\frac{\left|m_{2}s_{12}^{2}-m_{1}c_{12}^{2}\right|}{\left|
m_{2}s_{12}^{2}-m_{1}c_{12}^{2}\right|+m_{3}}.
\end{equation}
For $m_1=s^2_{12}\sqrt{\small{\frac{\Delta m_{12}^2}{ \cos 2
\theta_{12}}}}$ [cf. Eq. (14)], $s^2_{13}$ becomes zero for which
arbitrary values of $\beta$ are allowed. This is the point through
which all the curves corresponding to different values of $\beta$
pass in Fig. 1. Different points on the ($\alpha,m_1$) plane
correspond to different values of $s_{13}^2$. However, the CHOOZ
bound allows only a limited region on the ($\alpha ,m_{1}$) plane,
so we shall constrain the ($\alpha,m_{1}$) parameter space in the
light of the CHOOZ bound. However, the Majorana CP-violating phase
$\beta$ can not be constrained. Since, M is bounded from above
($\alpha =0^o$) and from below ($\alpha =90^o$) [cf. Eq. (22)],
$s_{13}^{2}$ is, also, bounded from above and below, i.e.
\begin{equation}
\frac{\left|m_{2}s_{12}^{2}-m_{1}c_{12}^{2}\right|}
{\left|m_{2}s_{12}^{2}-m_{1}c_{12}^{2}\right|+m_{3}}
<s_{13}^{2}<\frac{m_{1}c_{12}^{2}+m_{2}s_{12}^{2}}{%
m_{1}c_{12}^{2}+m_{2}s_{12}^{2}+m_{3}}.
\end{equation}
Thus, the region allowed by $m_{ee} =0$ on the
($m_{1},s_{13}^{2}$) plane will be bounded by the two limiting
values of $s_{13}^{2}$ corresponding to $\alpha=0^{o}$ and
$90^{o}$. As $m_{1}\rightarrow 0$, $s_{13}^2$ becomes independent
of $\alpha$ and is given by
\begin{equation}
s_{13}^{2}=\frac{\sqrt{\Delta m^2_{12}} s_{12}^{2}}{\sqrt{\Delta
m^2_{12}} s_{12}^{2}+\sqrt{\Delta m^2_{23}}}.
\end{equation}
This situation has been depicted in Fig. 2 where the upper and
lower bounds on $s_{13}^2$ [cf. Eq. (30)] have been plotted as
functions of $m_{1}$. The upper ($\alpha=0^o$) and lower
($\alpha=90^o$) curves come closer as $m_{1}\rightarrow 0$ and,
eventually, merge for [cf. Eq. (31)]
\begin{equation}
s^2_{13}=0.055^{+0.010,0.033}_{-0.009,0.024}.
\end{equation}
The central value of $s^2_{13}$ for $m_1=0$ given above is above
the CHOOZ bound but is consistent with the CHOOZ bound at about
2.3 $\sigma$. Therefore, a vanishing $m_{ee}$ element is
disallowed at 2.3 $\sigma$ C.L. for $m_{1}=0$ contrary to the
results reported in Ref. [3] where the special case $m_1=m_{ee}=0$
has also been examined and it has been found to be consistent with
the neutrino oscillation data. However, our results are consistent
with the recent results reported by Chauhan et al. \cite{4} who
conclude that the CHOOZ bound is violated when $m_1=m_{ee}=0$ in
vanishing determinant neutrino mass matrix scenarios where $m_1=0$
is a natural choice. As $m_1$ increases, the lower curve
(corresponding to $ \alpha=90^o$) dips to its minimum value
$s_{13}^2=0$ for $m_1=s^2_{12}\sqrt{\small{\frac{\Delta m_{12}^2}{
\cos 2 \theta_{12}}}}=0.045 eV$ [cf. Eq. (14)]. At this point,
$m_{ee}=0$ irrespective of the value of $\beta$. With further
increases in $m_{1}$, $s_{13}^2$ increases and, eventually,
becomes larger than the CHOOZ bound. Thus, there exists a range of
$m_{1}$ centered around 0.0045 eV and a corresponding range for
$\alpha$ around $90^o$, for which the condition $m_{ee}=0$ and the
CHOOZ bound hold simultaneously. Consequently, the  condition
$m_{ee}=0$ combined with the CHOOZ bound yields both an upper and
a lower bound on $m_1$. With further increase in $m_{1}$, we
approach the quasi-degenerate (QD) region for normal neutrino mass
ordering. For a certain value of $m_{1}$ ($\simeq 10^{-2}eV$ for
the central values of neutrino oscillation parameters given in Eq.
(10)), the value of $s_{13}^{2}$ becomes larger than the CHOOZ
bound. Thus, a vanishing $m_{ee}$ element is not allowed in QD
hierarchy with normal ordering of neutrino masses and, therefore,
a texture zero at the $ ee$ entry in the neutrino mass matrix in
the flavor basis can not accommodate a QD mass spectrum and small
$\theta _{13}$ simultaneously. When $m_1$ becomes very large as
compared to $\sqrt{\Delta m^2_{23}}$, Eq. (30) becomes
\begin{equation}
\frac{\left|s_{12}^{2}-c_{12}^{2}\right|}{\left|s_{12}^{2}-c_{12}^{2}\right|+1}
<s_{13}^{2}<\frac{1}{2}.
\end{equation}
The upper ($\alpha =0^{o}$) and lower ($\alpha =90^{o}$) curves in
Fig. 2 approach the upper and lower bounds on $s_{13}^{2}$ given
above asymptotically with $m_{1}$.

\begin{figure}[tb]
\begin{center}
\rotatebox{0}{\epsfig{file=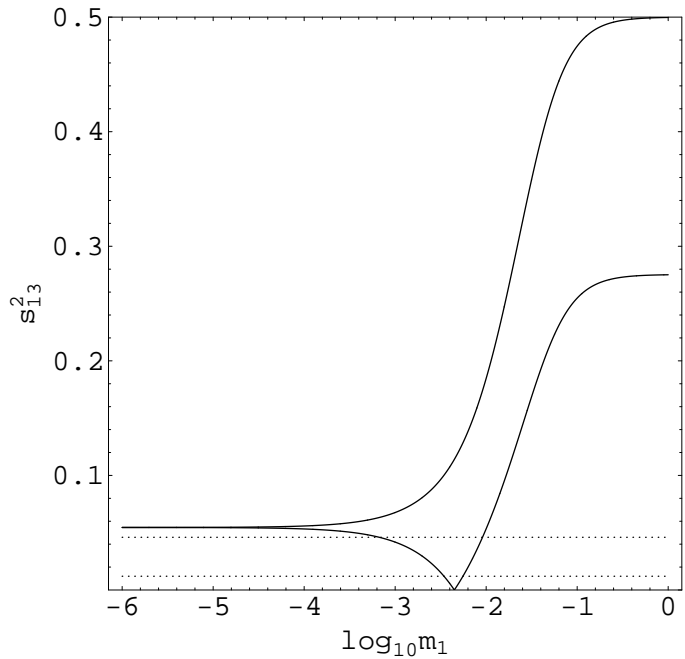, width=9.0cm, height=9.0cm}}
\end{center}
\caption{The ($s_{13}^2,m_{1}$) parameter space allowed by $
m_{ee} =0$ for $0^o<\alpha<90^o$. The upper solid line corresponds
to $\alpha=0^o$ and the lower solid line corresponds to
$\alpha=90^o$. We have also shown the 1 $\sigma$ and 3 $\sigma$
CHOOZ bounds as dotted lines.}
\end{figure}

\begin{figure}[tb]
\begin{center}
\rotatebox{0}{\epsfig{file=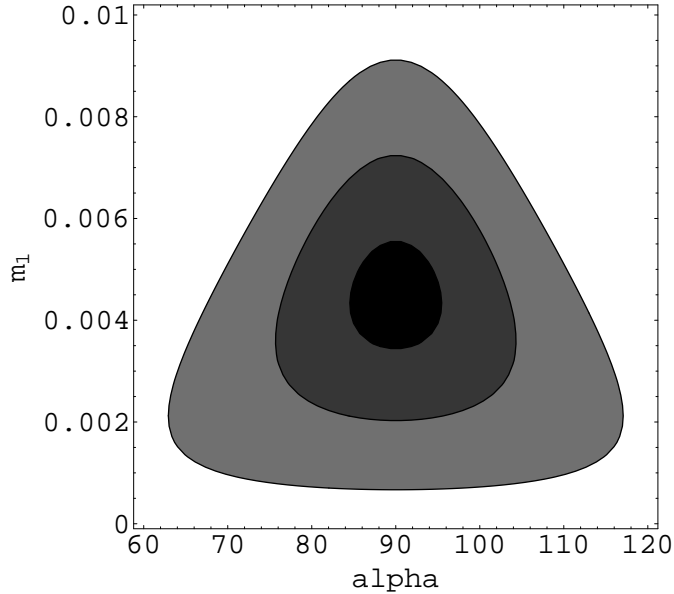, width=9.0cm, height=9.0cm}}
\end{center}
\caption{The parameter space on ($\alpha -m_{1}$) plane
constrained by $\left| m_{ee}\right| =0$ and the CHOOZ bound. The
regions in the increasing order of lighting correspond to 1
$\sigma$, 2 $\sigma$ and 3 $\sigma$ bounds on $s^2_{13}$}.
\end{figure}

\begin{table}[b]
\begin{center}
\begin{tabular}{|c|c|c|}
\hline C.L. & $m_1$ & $\alpha$  \\ \hline 1
$\sigma$&$0.0029-0.0063 eV$ & $83.9^o-96.1^o$ \\ 2
$\sigma$&$0.0011-0.0093 eV$ & $71.2^o-108.8^o$   \\ 3
$\sigma$&$0-0.0129eV$ & $33.7^o-146.3^o$
\\ \hline
\end{tabular}
\end{center}
\caption{ The allowed values of $m_1$ and $\alpha$ for $m_{ee}=0$
from the neutrino oscillation data at various confidence levels.}
\end{table}

In Fig. 3, the contours for 1 $\sigma$, 2 $\sigma$ and 3 $\sigma$
values of $s^2_{13}$ allowed by the CHOOZ bound have been plotted
on the ($\alpha,m_{1}$) plane. The allowed range of $m_1$ in Fig.
3 is essentially the same as in Fig. 2. The values of $\alpha$
which give $m_{ee}=0$ are centered around $\alpha=90^o$ and the
($\alpha,m_{1}$) parameter space becomes smaller with the decrease
in the upper bound on $s^2_{13}$. At two standard deviations, the
allowed range of $\alpha$ is $75^o-105^o$ approximately and the
best fit point on the ($\alpha,m_{1}$) plane is approximately
$(90^o,0.0045eV)$. The exact value of $m_1$ for which $m_{ee}$
vanishes is given by Eq. (14) and this happens for $s_{13}=0$ and
$\alpha=90^o$.

While plotting Fig. 2 and Fig. 3, we used the central values of
the neutrino oscillation parameters. Now, we incorporate the
experimental errors in these parameters in the numerical analysis
to fix the allowed ranges of $m_{1}$ and $\alpha$. The results of
this analysis have been summarized in Table 1. The allowed ranges
for $m_1$ and $\alpha$ shown in Fig. 3, however, come from the 1
$\sigma$, 2 $\sigma$ and 3 $\sigma$ ranges for $s^2_{13}$ and
central values for the other oscillation  parameters and,
therefore, are more restrictive as compared to the ranges given in
Table 1.

Thus, the effective Majorana mass can be zero in normal hierarchy
if $m_1$ and $\alpha$ lie within the ranges shown in Table 1. The
inter-relationship between the hierarchy of the neutrino mass
spectrum and the resulting texture structure of the neutrino mass
matrix is apparent from this analysis. The mass matrices with a
texture zero at the $ee$ entry naturally lead to a mass spectrum
with normal hierarchy and small mixing angle $\theta_{13}$.

\begin{figure}[tb]
\begin{center}
\rotatebox{0}{\epsfig{file=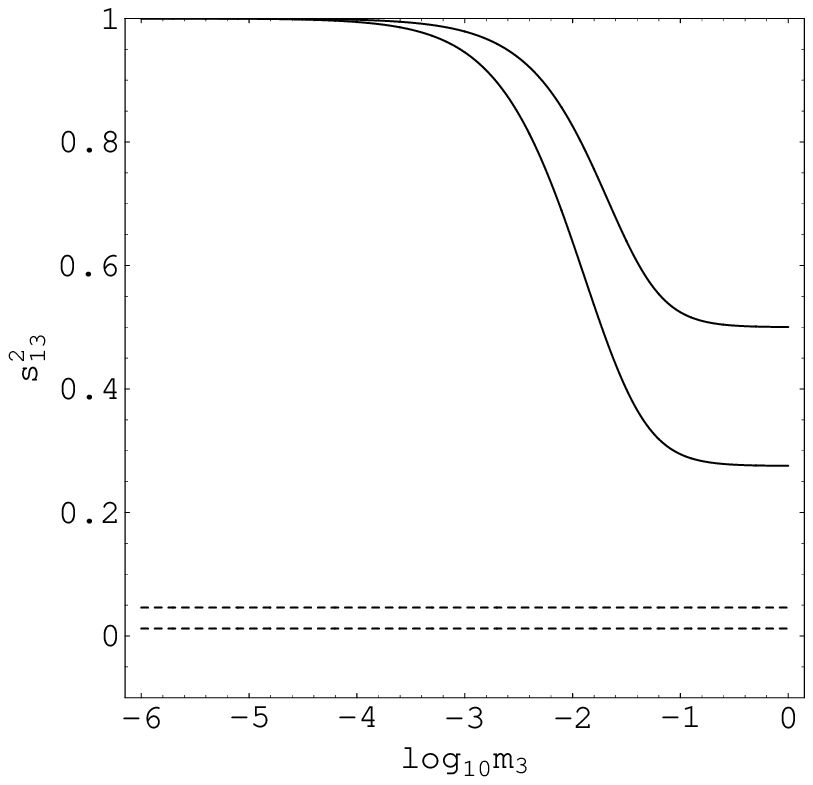, width=9.0cm, height=9.0cm}}
\end{center}
\caption{The parameter space allowed on ($s^2_{13}-m_{1}$) plane
by $\left| m_{ee}\right| =0$ for $0^o<\alpha<90^o$ for the
inverted hierarchy. The upper solid line corresponds to
$\alpha=0^o$ and the lower solid line corresponds to
$\alpha=90^o$. We have also shown the CHOOZ 1 $\sigma$ and 3
$\sigma$ bounds as dotted lines.}
\end{figure}

Finally, we examine the case of inverted hierarchy where the
lightest neutrino mass is $m_3$. It is clear from Eq. (25) that
$s_{13}^2=1$ for $m_3=0$. As $m_3$ increases, $s_{13}^2$ decreases
and attains its minimum value asymptotically. This feature is
apparent from Fig. 4 where $s_{13}$ has been plotted as a function
of $m_3$ for inverted hierarchy. For $m_{3}$ very large, we obtain
\begin{equation}
\frac{\left|s_{12}^{2}-c_{12}^{2}\right|}{\left|s_{12}^{2}-c_{12}^{2}\right|+1}
<s_{13}^{2}<\frac{1}{2}.
\end{equation}
from Eq. (30). The lower and upper bounds on $s_{13}^2$ given
above correspond to the asymptotic values of the upper
($\alpha=0^0$) and lower ($\alpha=90^o$) curves in Fig. 4
respectively. It is easily seen that the CHOOZ bound is much below
the values of $s_{13}^2$ required for vanishing $m_{ee}$ in
inverted hierarchy as well as in quasi-degenerate hierarchy with
the inverted ordering of neutrino masses. Therefore, neutrino mass
matrices with a texture zero at the $ee$ entry cannot yield a mass
spectrum with inverted hierarchy and small $\theta_{13}$
simultaneously. This happens because the first mass eigenvalue
will be smaller than the second and third eigenvalues if the $ee$
element of the mass matrix is zero unless a very large 1-3
rotation is effected to make it larger than the third mass
eigenvalue.

In conclusion, the consequences of a vanishing effective Majorana
mass have been examined in detail. It has been concluded that the
effective Majorana mass can be zero only for normal hierarchy for
certain ranges of values of $m_1$ and $\alpha$. The Majorana phase
$\beta$ is left unconstrained but it should be one quadrant behind
$\alpha$. It is found that effective Majorana mass is not allowed
to vanish at 2.3 $\sigma$ C.L. when $m_1=0$ in normal hierarchy.
Mass matrices with a texture zero at the $ee$ entry naturally lead
to a normal ordered neutrino mass spectrum and small
$\theta_{13}$. However, a neutrino mass spectrum with inverted or
even quasi-degenerate hierarchies is not allowed by the condition
of a vanishing effective Majorana mass in the light of the current
neutrino data.

\bigskip

\textit{\Large{Acknowledgements}}

The research work of S. D. is supported by the Board of Research
in Nuclear Sciences (BRNS), Department of Atomic Energy,
Government of India \textit{vide} Grant No. 2004/ 37/ 23/ BRNS/
399. S. K. acknowledges the financial support provided by Council
for Scientific and Industrial Research (CSIR), Government of
India.

\end{document}